\documentclass[11pt,english]{article}
\usepackage{ae,aecompl}

\usepackage[T1]{fontenc}
\usepackage[latin9]{inputenc}
\usepackage{geometry}
\usepackage{amsmath,amssymb}
\usepackage{esint}
\usepackage{graphicx}
\usepackage{subfigure}
\usepackage{float}
\usepackage{color}
\usepackage{cite} 
\usepackage{fullpage}
\usepackage{hyperref} 
\usepackage{pdflscape}
\usepackage{footnote}
\usepackage{caption}
\usepackage{mathrsfs} 
\usepackage{multirow}

\newcommand{\be}{\begin{equation}}
\newcommand{\ee}{\end{equation}}
\newcommand{\beq}{\begin{equation}}
\newcommand{\eeq}{\end{equation}}
\newcommand{\bea}{\begin{eqnarray}}
\newcommand{\eea}{\end{eqnarray}}
\newcommand{\besp}{\begin{equation}\begin{split}}
\newcommand{\eesp}{\end{split}\end{equation}}

\newcommand{\Br}{\text{Br}}

\newcommand{\Eq}[1]{Eq.~(\ref{#1})}

\def\mL{\mathcal{L}}

\makeatletter
\usepackage{slashed}

\makeatother

\usepackage{babel}

\begin{document}

\title{Scalar-mediated dark matter model at colliders and gravitational wave detectors -- A White paper for Snowmass 2021}

\date{}

\author{Jia Liu$^{a,b}$, Xiao-Ping Wang$^{c}$ and Ke-Pan Xie$^{d}$}

\maketitle

\begin{center}
\it {$^a$ School of Physics and State Key Laboratory of Nuclear Physics and Technology, Peking University, Beijing 100871, China}

\it {$^b$ Center for High Energy Physics, Peking University, Beijing 100871, China}

\it {$^c$ School of Physics and Nuclear Energy Engineering, Beihang University, Beijing 100083, China}

\it {$^d$ Department of Physics and Astronomy, University of Nebraska, Lincoln, NE 68588, USA}
\end{center}

\begin{abstract}
The weakly interacting massive particles (WIMPs) have been the most popular particle dark matter (DM) candidate for the last several decades, and it is well known that WIMP can be probed via the direct, indirect and collider experiments. However, the direct and indirect signals are highly suppressed in some scalar-mediated DM models, e.g. the lepton portal model with a Majorana DM candidate. As a result, collider searches are considered as the only hope to probe such models. In this white paper, we propose that the gravitational wave (GW) astronomy also serves as a powerful tool to probe such scalar mediated WIMP models via the potential first-order phase transition GW signals. An example for the lepton portal dark matter is provided, showing the complementarity between collider and GW probes.
\end{abstract}

\section{Motivation}

Although dark matter (DM) contributes as large as $\sim85\%$ to the matter of the Universe~\cite{Planck:2018vyg}, its origin remains as a long-standing mystery in particle physics~\cite{Bertone:2004pz}. Over the past several decades, the weakly interacting massive particle (WIMP) paradigm~\cite{Lee:1977ua} has been the most popular explanation for particle DM. The ``WIMP miracle'' allows a natural explanation for electroweak scale DM particles and interactions. What's more, the WIMP models can be typically probed via the direct~\cite{Schumann:2019eaa}, indirect~\cite{Gaskins:2016cha} and collider~\cite{Boveia:2018yeb} searches. However, in many WIMP models, the direct and indirect signals are accidentally suppressed that collider searches are usually expected as the only hope to probe the DM parameter space.

One such example is the lepton portal DM model with Majorana DM candidate~\cite{Bai:2014osa} (also see Refs.~\cite{Yu:2014mfa, Altmannshofer:2014cla, Yu:2014pra, Garny:2015wea, Agrawal:2015tfa, Ibarra:2015fqa, Cai:2015zza, Baek:2015fma, Chen:2015jkt, Berlin:2015njh, Agrawal:2015kje,  Mukherjee:2015axj, Evans:2016zau, Chao:2016lqd, Borah:2017dqx, Kowalska:2017iqv, Duan:2017pkq,  Yuan:2017ysv, Tang:2017lfb, Ge:2017tkd, Ding:2017jdr, Baker:2018uox, Hisano:2018bpz, Gaviria:2018cwb, Kavanagh:2018xeh, Kawamura:2020qxo, Okada:2020oxh, Ge:2020tdh, Boehm:2020wbt, Okawa:2020jea, Kowalska:2020zve, Verma:2021koo, Alvarado:2021fbw, Horigome:2021qof, Bai:2021bau,Jueid:2020yfj,Arcadi:2021glq,Arcadi:2021cwg,Calibbi:2018rzv}). In this model, the DM candidate $\chi$ couples to the Standard Model (SM) particles via a complex scalar $S$ which is charged $-1$ under the hypercharge group. This model has negligible direct and indirect search signals: the nuclear recoil cross section in direct detection comes from loop diagrams; while the annihilation cross section suffers from helicity suppression~\cite{Bai:2014osa}. Therefore, collider experiments are crucial in probing the model.\footnote{See Refs.~\cite{Cermeno:2022rni,Cermeno:2021rtk} for the gamma ray signals for this model.} In Ref.~\cite{Liu:2021mhn}, we have studied the hadron and electron-positron collider phenomenology of this model, and pointed out that the first-order electroweak phase transition gravitational wave (GW) signals can also probe a considerable fraction of the parameter space. The idea in that paper can be generalized into other WIMP models which are difficult to probe via the direct and indirect experiments.

\section{Probing the lepton portal DM model with collider and GWs}

In this section, we summarize the study in Ref.~\cite{Liu:2021mhn}. The relevant Lagrangian for the lepton portal DM with a Majorana DM candidate is
\begin{align}\label{eq:Lags}
\mL_\chi =&~ \frac12\bar\chi i\slashed{\partial}\chi-\frac12m_\chi\bar\chi\chi+ y_\ell\left(\bar{\chi}_L S^\dagger \ell_R+{\rm H.c.}\right), \\
\mL_S =&~ \left(D^\mu S \right)^\dagger D_\mu S - V(H,S),\\
V(H, S)=&~\mu_H^2|H|^2+\mu_S ^2| S |^2+\lambda_H|H|^4+\lambda_S | S |^4+2\lambda_{H S }|H|^2| S |^2,
\end{align}
where $\chi$ is the DM candidate (gauge singlet), while $S$ is the mediator ($SU(2)_L$ singlet but charged $-1$ under $U(1)_Y$). $H$ and $\ell$ are respectively the SM Higgs doublet and lepton ($\ell=e$, $\mu$, $\tau$). The annihilation $\chi\chi\to\ell^+\ell^-$ via the exchange of a $t$-channel mediator dominates the DM relic abundance after freeze-out. Although both the direct and indirect signals are suppressed, the model can be probed via collider signals.

The lepton portal coupling $y_\ell$ and scalar portal coupling $\lambda_{HS}$ can be efficiently probed by the following channels.
\begin{enumerate}
\item The pair production and exclusive decays $S^+S^-\to\ell^+\chi\ell^-\chi$, leading to a di-lepton plus missing energy final state. At the LHC, the combination of Drell-Yan $q\bar q\to Z^*/\gamma^*\to S^+S^-$ channel and the gluon-gluon fusion channel $gg\to h^*\to S^+S^-$ can probe the scalar portal coupling $\lambda_{HS}$; while at the future lepton collider such as FCC-ee and CEPC, the off-shell production of $e^+e^-\to Z^*/\gamma^*\to S^{\pm (*)} S^{\mp}\to\ell^+\chi\ell'^-\chi$ offers the opportunity to probe $y_\ell$ directly.

\item Exotic decays of the Higgs or $Z$ boson. The first case is the exotic decay to a pair of leptons and missing energy $h/Z\to S^{\pm (*)} S^{\mp (*)}\to\ell^+\chi\ell'^-\chi$. In the Higgs case, the decay width is $\propto y_\ell^2\lambda_{HS}^2$ or $\propto y_\ell^4\lambda_{HS}^2$, providing a new way to probe $\lambda_{HS}$ and $y_\ell$; while in the $Z$ case, the width is $\propto y_\ell^4$. The second case is the Higgs invisible decay $h\to \chi\chi$, which can probe the combination $y_\ell^2\lambda_{HS}$.

\item One loop corrections to the Higgs couplings, including $h\ell^+\ell^-$, $h\gamma\gamma$ and $hZZ$. The leptonic coupling probes the combination $y_\ell^2\lambda_{HS}$, while the $h\gamma\gamma$ and $hZZ$ couplings probe $\lambda_{HS}$ solely.

\item The electron or muon anomalous magnetic moment. Unfortunately, in the minimal model considered in \Eq{eq:Lags}, the contribution is negative that it can not explain the muon $g-2$ anomaly~\cite{Abi:2021gix}. While in the future study, if there are pseudo scalars' positive contribution to the muon magnetic moment, we probably can address the muon $g-2$ anomaly. On the other hand, if Yukawa interactions of scalar mediator induce the effective photon interactions, two loop effects will also need to be considered for muon $g-2$. 
\end{enumerate}

\begin{figure}
\centering	
\includegraphics[width=0.4 \columnwidth]{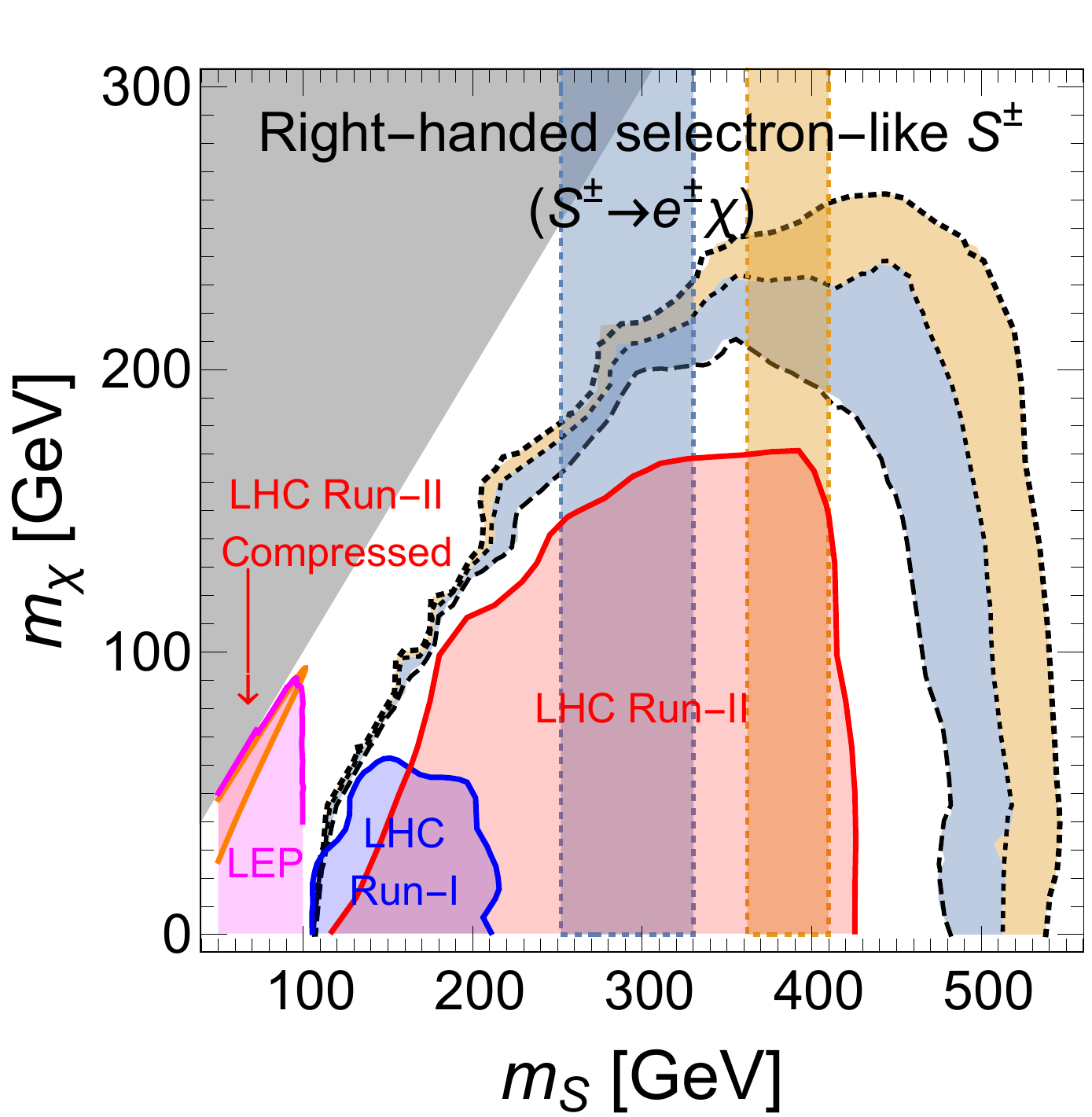} \qquad
\includegraphics[width=0.4 \columnwidth]{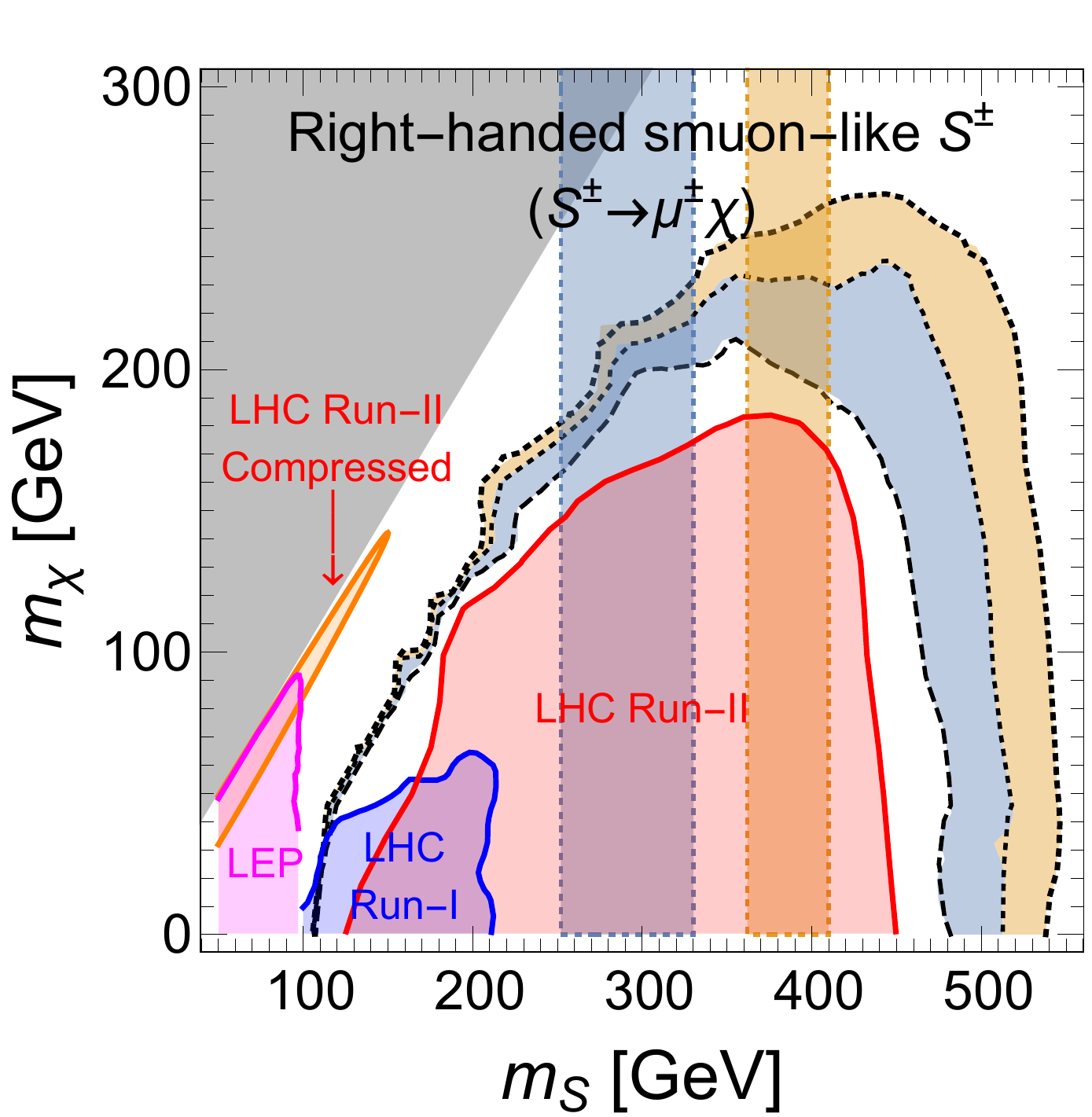}
\caption{Figure from Ref.~\cite{Liu:2021mhn}. The interplay between gravitational wave detection and LHC searches. The shaded regions are exclusions from LEP~\cite{LEPslepton}, LHC Run-I ($20.3~{\rm fb}^{-1}$)~\cite{Aad:2014vma} and LHC Run-II ($139~{\rm fb}^{-1}$)~\cite{Aad:2019vnb,Aad:2019qnd}.	The black dashed lines are projections for the LHC reach at 300 ${\rm fb}^{-1}$ with only the Drell-Yan production of $S^+S^-$. The light blue (orange) shaded region corresponds to $\lambda_{HS}=2$ (3), with the vertical boxed boundary regions being the LISA-detectable parameter space, while the irregular boundary regions being enhanced part of the LHC projections when including the $gg\to h^*\to S^+S^-$ contribution.}
\label{fig:ms-mchi-GW}
\end{figure}

Besides the collider signals, the model might also trigger a first-order phase transition in the early Universe, provided that the mediator bare mass term $\mu_S^2$ is negative, and the portal coupling $\lambda_{HS}$ is large enough. Therefore, the phase transition GWs can also be a probed for $\lambda_{HS}$. The interplay between LHC searches and GW probes are shown in Fig.~\ref{fig:ms-mchi-GW}, where the LHC and LISA projections are both shown in the $m_S$-$m_\chi$ plane. The $pp\to S^+S^-\to\ell^+\chi\ell^-\chi$ reach can be enhanced if $\lambda_{HS}\neq 0$ such that the off-shell Higgs mediated production $gg\to h^*\to S^+S^-$ also plays a role. On the other hand, a large enough $\lambda_{HS}$ can trigger the first-order phase transition and leave signals at the GW detectors like LISA, Taiji, TianQin, BBO and U-DECIGO. The enhancement of LHC reach and the LISA projections are shown in light blue and orange regions for $\lambda_{HS}=2$ and 3, respectively. One can see that the LHC and LISA experiments mainly serve as complementary approaches to probe the DM parameter space; while they also have some intersections, which can be used to identify the origin of the excess detected in the future.

\begin{figure}
\centering	
\includegraphics[width=0.32 \columnwidth]{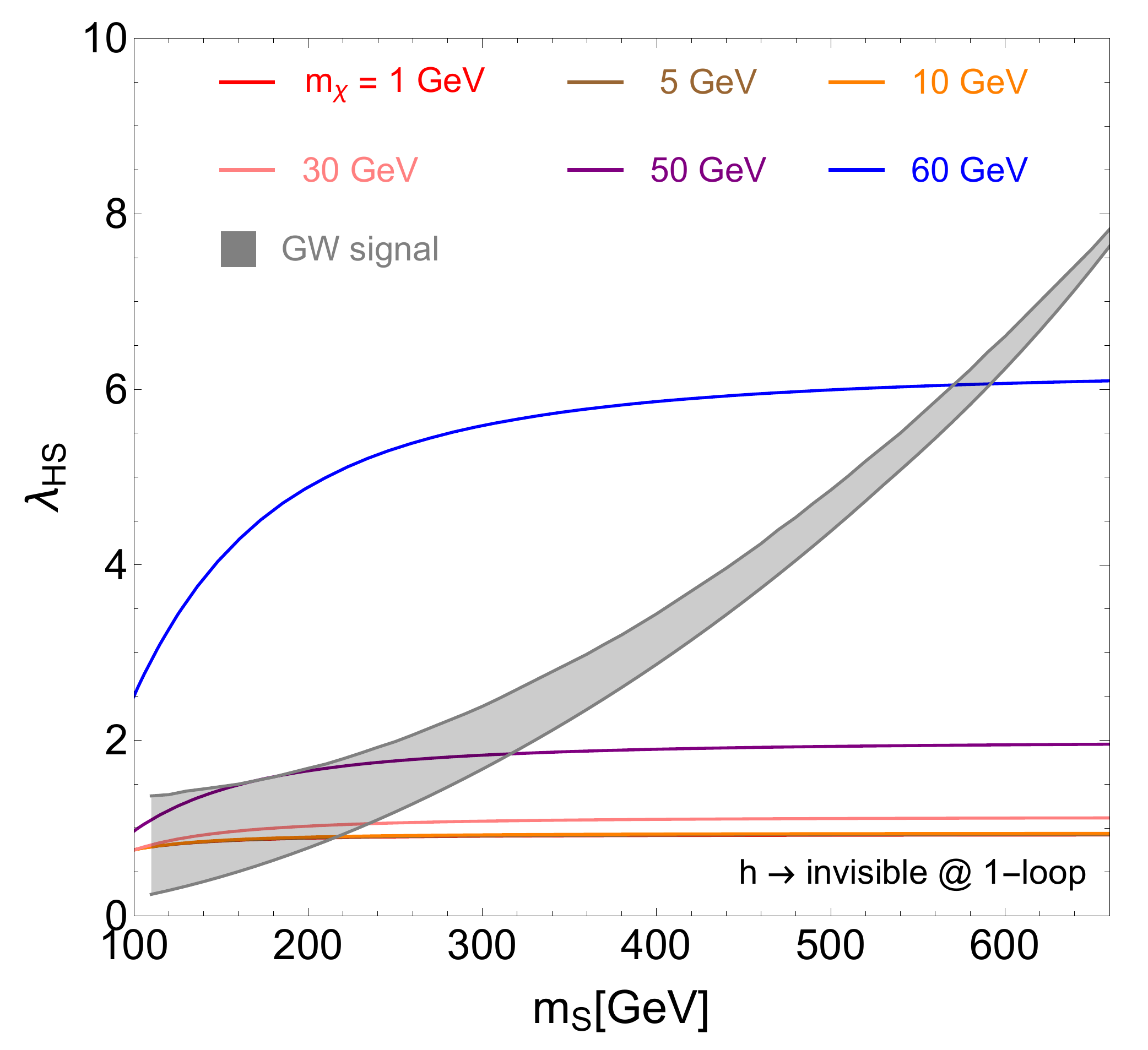}
\includegraphics[width=0.32 \columnwidth]{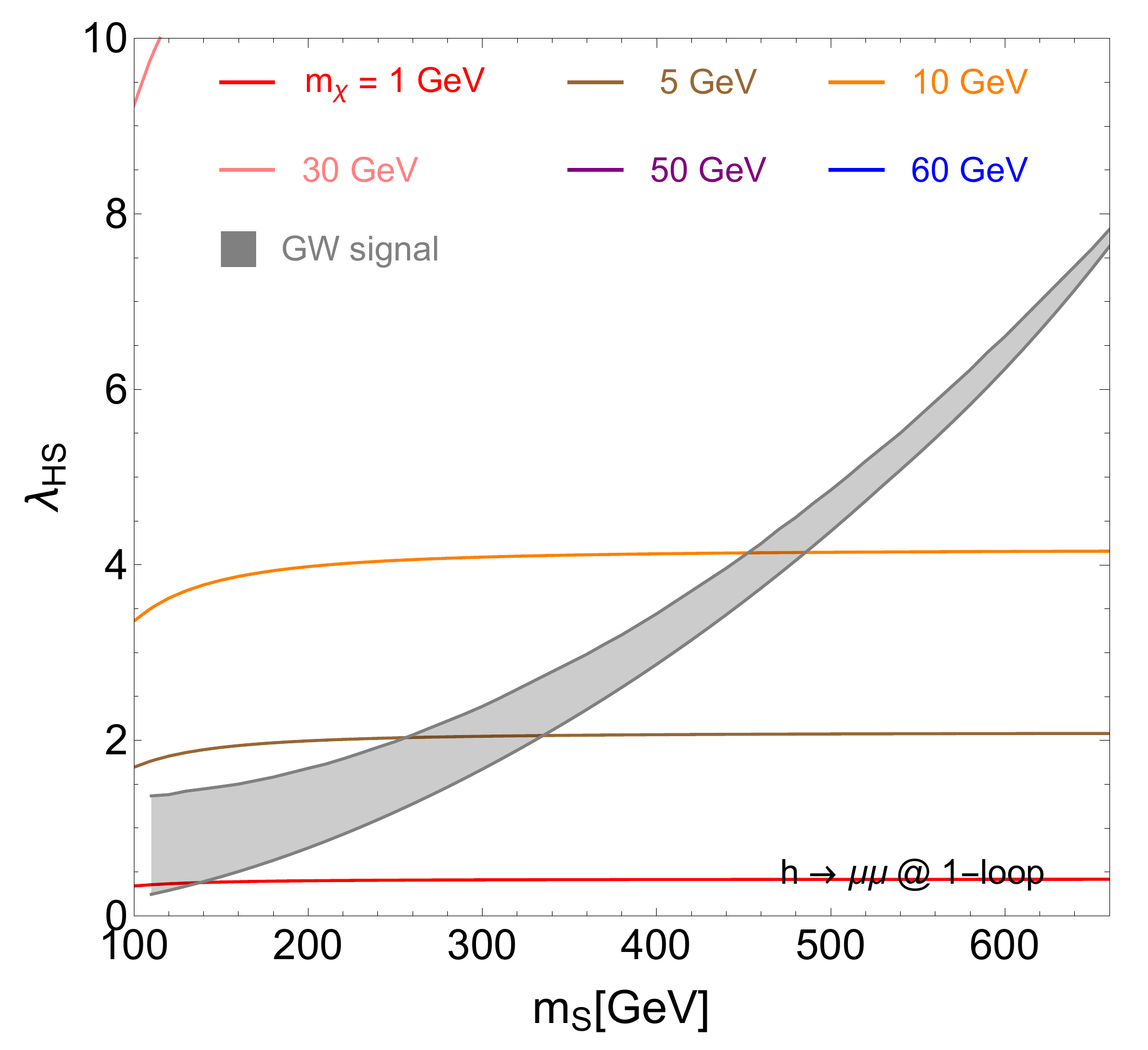}
\includegraphics[width=0.32 \columnwidth]{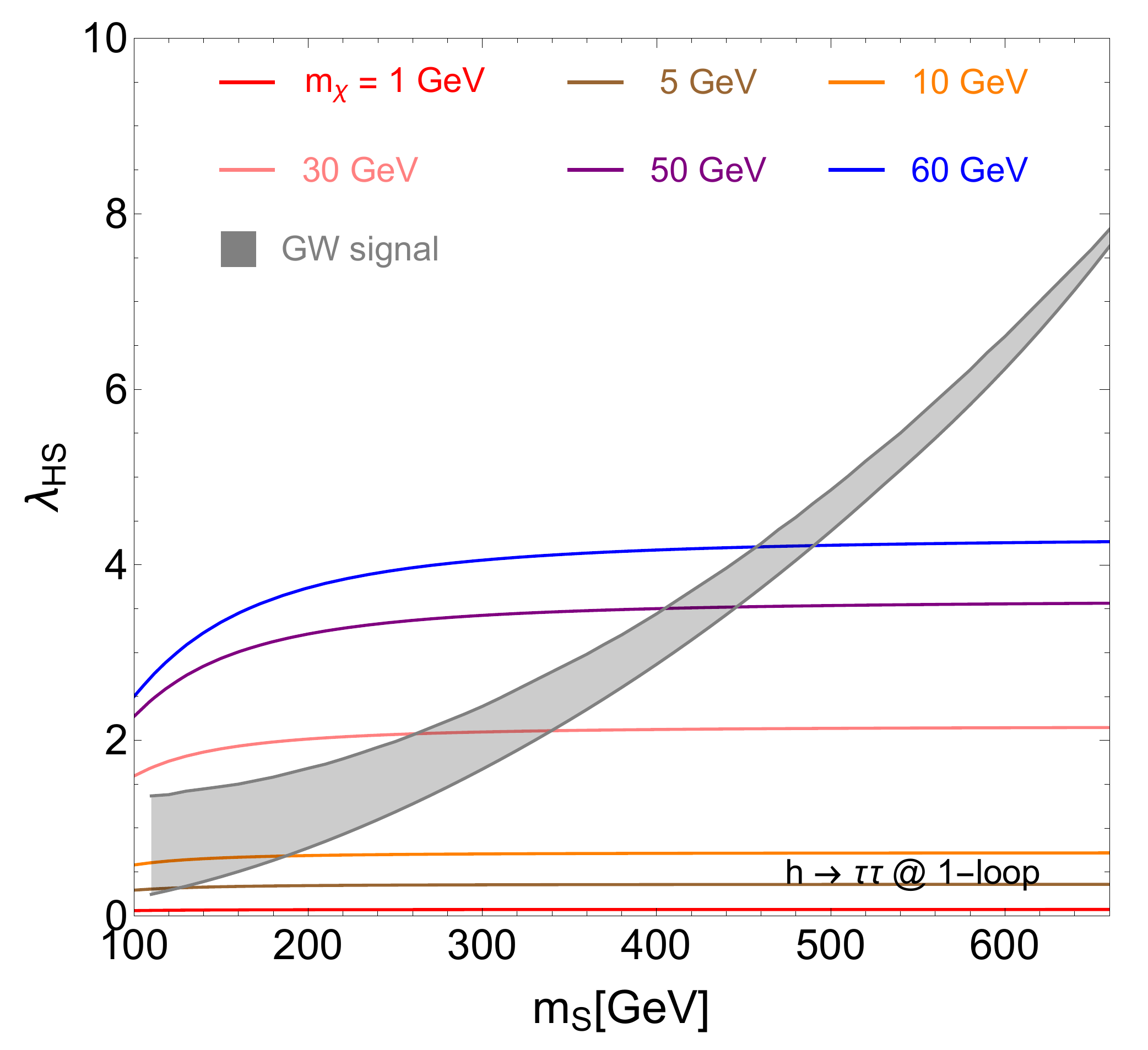}
\caption{Figure from Ref.~\cite{Liu:2021mhn}. The interplay between GW detection and future $e^+e^-$ collider searches. The gray shaded region is the LISA detectable parameter space, varying $\lambda_S$ from 0 to $4 \pi$. From left to right, we show the sensitivities for $\lambda_{HS}$ from future FCC-ee and CEPC precision measurements, based on invisible Higgs decay branching ratio $\Br(h\to{\rm inv}) = 0.3\%$, Higgs leptonic coupling precision reaches $\delta \kappa_\mu < 8.7\%$ and $\delta \kappa_\tau <1.5\%$.}
\label{fig:h-1loop-GW}
\end{figure}

The interplay between GW detections and future $e^+e^-$ collider searches are shown in Fig.~\ref{fig:h-1loop-GW}, where we show the LISA projections and the CEPC Higgs measurement projections for comparison. The first-order phase transition and LISA-detectable GW parameter space are determined by $m_S$ and $\lambda_{HS}$, as shown in gray shaded regions. The CEPC projections for Higgs invisible decay, $h\mu\mu$ and $h\tau\tau$ couplings depend an extra parameter, the DM mass $m_\chi$. Given an $m_\chi$, the region above the colored lines in Fig.~\ref{fig:h-1loop-GW} can be probed by the CEPC with a collision energy of 240 GeV and an integrated luminosity of 5 ab$^{-1}$. The intersections between the LISA and CEPC projections can be used for crosschecking the excess obtained in either approach; while in other parameter space the two approaches are complementary.

\section{Summary and outlook}

We have shown in Ref.~\cite{Liu:2021mhn} that collider searches can probe the lepton portal DM model efficiently, and the GW detection experiments can be an important complementary crosscheck. In general, such complementarity between collider searches and GW experiments exist in many WIMP models with scalar DM and/or scalar mediators, dubbed dark scalars $S$, which are usually oddly charged under a $\mathbb{Z}_2$ symmetry. This is because the joint potential between Higgs and the dark scalars, especially the Higgs portal coupling $|S|^2|H|^2$, is inevitable in the model, and the Higgs portal coupling could serve as the source of the potential barrier that trigger the first-order phase transition. Therefore, the WIMP model might manifest itself both via collider and GW signals in the future, and a correlation/complementarity study would be extremely useful. 

In the future, more studies can be done along this line, for example in Ref.~\cite{Liu:2021mhn} the charged singlet scalar mediator is considered while doublet scalar as mediator is also possible. Moreover, the study shows the example for slepton-like scalars and one can extend to squark-like scalar as mediator as well. In addition, DM itself as scalar can also be possible to show its relevance at colliders and GW probes. In this aspect, single scalar DM model has been extensively studied already~\cite{Jiang:2015cwa,Shajiee:2018jdq,Han:2020ekm,Zhang:2021alu,Borah:2021ocu,Costa:2022oaa}, while higher multiplets scalar DM is less studied and worth probing.

\section*{Acknowledgments}
The work of J.L. is supported by National Science Foundation of China under Grant No. 12075005
and by Peking University under startup Grant No. 7101502597.
The work of X.P.W. is supported by National Science Foundation of China under Grant No. 12005009.
K.P.X. is supported by the University of Nebraska-Lincoln.

\bibliographystyle{utphys}
\bibliography{references}
\end{document}